\documentclass[reprint,pra,superscriptaddress,nofootinbib,aps,twocolumn]{revtex4-2}
\usepackage{amsmath, amssymb}	
\usepackage{bbm}				
\usepackage{braket}				
\usepackage{graphicx} 			
\usepackage{mathtools}			
\usepackage{hyperref}			
\usepackage{cleveref}
\usepackage{dsfont}
\usepackage{bm}
\usepackage[T1]{fontenc}
\usepackage{mathrsfs}
\usepackage{xcolor}
\usepackage[section]{placeins}

\begin{document}
\title{Information scrambling and entanglement in quantum approximate optimization algorithm circuits}

\author{Chen Qian}
\email[]{qianchen@baqis.ac.cn}
\affiliation{Beijing Academy of Quantum Information Sciences, Beijing 100193, China}
\affiliation{State Key Laboratory of Low Dimensional Quantum Physics, Department of Physics, Tsinghua University, Beijing, 100084, China}

\author{Wei-Feng Zhuang}
\affiliation{Beijing Academy of Quantum Information Sciences, Beijing 100193, China}

\author{Rui-Cheng Guo}
\affiliation{State Key Laboratory of Low Dimensional Quantum Physics, Department of Physics, Tsinghua University, Beijing, 100084, China}

\author{Meng-Jun Hu}
\affiliation{Beijing Academy of Quantum Information Sciences, Beijing 100193, China}

\author{Dong E. Liu}
\email[ ]{dongeliu@mail.tsinghua.edu.cn}
\affiliation{Beijing Academy of Quantum Information Sciences, Beijing 100193, China}
\affiliation{State Key Laboratory of Low Dimensional Quantum Physics, Department of Physics, Tsinghua University, Beijing, 100084, China}
\affiliation{Frontier Science Center for Quantum Information, Beijing 100184, China}

\begin{abstract}
Variational quantum algorithms, which consist of optimal parameterized quantum circuits, are promising for demonstrating quantum advantages in the noisy intermediate-scale quantum (NISQ) era. Apart from classical computational resources, different kinds of quantum resources have their contributions in the process of computing, such as information scrambling and entanglement. Characterizing the relation between complexity of specific problems and quantum resources consumed by solving these problems is helpful for us to understand the structure of VQAs in the context of quantum information processing. In this work, we focus on the quantum approximate optimization algorithm (QAOA), which aims to solve combinatorial optimization problems. We study information scrambling and entanglement in QAOA circuits respectively, and discover that for a harder problem, more quantum resource is required for the QAOA circuit to obtain the solution in most of the cases. We note that in the future, our results can be used to benchmark complexity of quantum many-body problems by information scrambling or entanglement accumulation in the computing process.   
\end{abstract}

\maketitle

\section{Introduction}\label{sec:introduction}
Quantum computation is considered to have a crucial computational speed-up compared with classical computing, because it consumes not only classical computational resources but also quantum resources~\cite{Chitambar-rmp-2019}. As the most important feature of quantum mechanics, entanglement offers essential resource for quantum computation~\cite{Horodecki-rmp-2007,Preskill-2018}. Besides, since quantum circuits are actually unitary channels, the quantity characterizing spatiotemporal entanglement properties of a channel, namely information scrambling, should be viewed as a considerable resource as well~\cite{Hosur-jhep-2016,Landsman-nat-2019,Xu-2202,Ahmadi-2204,Garcia-2208}. The concept of information scrambling was first introduced by Hayden and Preskill~\cite{Hayden-jhep-2007} to demonstrate that the quantum information of a diary thrown into a black hole will spread to the whole system, and finally recover from the black hole evaporation. In last decades, studies of this physical process mostly focus on the area of black hole information and quantum gravity~\cite{Harlow-rmp-2016,Yoshida-2019}. Currently, it has been demonstrated that for the H-P protocol, scrambled information can be efficiently decoded using a Clifford decoder, provided that the system is not fully chaotic~\cite{Oliviero-2212,Leone-2212}. Recently, more and more works turn their attention to information scrambling in quantum circuits, which pave the way to applications involving benchmarking noise~\cite{Harris-prl-2022}, recovering lost information~\cite{Yan-prl-2020}, characterizing performance of quantum neural networks~\cite{Shen-prl-2020,Wu-prr-2021,Garcia-jhep-2022}, unifying chaos and random circuits~\cite{Cotler-jhep-2017,Roberts-jhep-2017}, and so on~\cite{Bhattacharyya-epjc-2022,Iyoda-pra-2018}. In addition, the process of information scrambling is observed experimentally on superconducting quantum processors~\cite{Mi-science-2021,Zhu-prl-2022}. Briefly, studying information scrambling and entanglement properties offers us an overview of how much quantum resource is needed for specific quantum circuits.

Variational quantum algorithms (VQAs) belong to the leading strategies to potentially present quantum advantage on near-term NISQ devices~\cite{Landsman-nat-2019}, which include a series
of algorithms such as variational quantum eigensolver (VQE)~\cite{Kandala-nat-2017}, quantum machine learning (QML)~\cite{Biamonte-nat-2017} and quantum approximate optimization algorithm (QAOA)~\cite{Farhi-1411}. VQAs share the common structure of optimizing parameters with a classical optimizer and running parameterized quantum circuits to solve problems. In particular, QAOA is proposed to solve the quadratic unconstrained binary optimization (QUBO) problems~\cite{QUBO_2022}, which can be translated into computing minimal expectation values of an Ising Hamiltonian. Based on these settings, QAOA is widely applied in combinatorial optimization problems, e.g., traffic congestion, finance, and many-body physics~\cite{Wu-prl-2019,Fitzek-2110}. It is useful to give a thorough analysis to the scrambling and entanglement properties of the optimal parameterized quantum circuits in order to benchmark the quantum resource consumed when we utilize VQAs to calculate a certain problem.

The role of entanglement for VQAs has been extensively studied from different perspectives~\cite{Wiersema-prxq-2020,Valle-pra-2021,Chen-2205,Dupont-pra-2022,Sreedhar-2207}. In recent papers, bipartite entanglement entropy in $p$-layer QAOA circuits was investigated. Chen, \emph{et al.} compared the entanglement required between ADAPT-QAOA and standard QAOA solving certain problems~\cite{Chen-2205}, and Dupont, \emph{et al.} characterize entanglement generated in QAOA circuit with entanglement volume law~\cite{Dupont-pra-2022}. However, details about entanglement generation in the $p$-layer circuit and its connection to complexity of problems still remains an open question. Moreover, in the aspect of information scrambling, Shen, \emph{et al.} established a correlation between scrambling ability and loss function in the training process of quantum neural networks~\cite{Shen-prl-2020}, which raises an interesting question of whether a comparable correlation occurs in QAOA.

Therefore, our aim in this paper is to investigate the connection between complexity of QUBO problems and quantum resource (including information scrambling and entanglement) consumed in the QAOA circuits. In this work, we address a special type of QUBO problems whose solutions are non-degenerate in order to concentrate on the quantum resource generated during the computing process. When we embed these solutions to quantum circuits, the target states are product states. Since both initial and final states have no entanglement, all quantum resource generated in the circuit is only for computing steps. As we have fixed all basic settings of the optimization algorithm, we can build a clear link between the amount of quantum resources used in QAOA circuits and the complexity to solve QUBO problems.

Next, we need to discriminate the concept ``complexity of QUBO problems''. This concept includes two different definitions~\cite{QUBO_2022}: 

1.~The complexity to solve a QUBO problem mathematically, which is highly dependent on structure of the graph.

2.~The computational difficulty of solving a QUBO problem using certain algorithms. Both algorithm and properties of the graph are relevant to this computational complexity.

The two complexities may not have a definitive correlation. As we have discussed, the mathematical complexity of a QUBO problem depends on its geometric structure, such as edges and weights. Usually, a mathematically harder problem is more difficult to solve using algorithms on a classical or quantum computer as well. However, for different algorithms, the difficulty to solve these QUBO problems may be different. In this work, we only discuss solving QUBO problems via QAOA, thus we take the second definition to ``complexity of QUBO problems'' throughout this paper. Although a general mathematical expression of computational complexity of solving QUBO problems via QAOA is still an open question, at least it is determined by some important attributes of a QUBO graph, like
\begin{equation}
\mathcal{C} \sim f\left(D,w_{ij},h_{k}\right),\label{eq:def-complexity}
\end{equation}
where $D$ is the density of a graph, $w_{ij}$ and $h_{k}$ are weights of edges and nodes of the graph, detailed definitions are introduced in Sec.~\ref{subsec:QAOA}.

Based on all above settings, we present our results in two parts: Firstly, we treat the QAOA circuit as a quantum channel to compare information scrambling characterized by tripartite information with QAOA computational complexity characterized by success rates, and our results are obtained for three different kinds of QUBO problems. Secondly, we compare entanglement accumulation in QAOA circuit with five QUBO problems where the corresponding graphs have different edges but the weights of edges and nodes are the same. Notably, we propose a quantity to characterize entanglement accumulation inside QAOA circuit, which has the form of average area of entanglement entropy generated during the evolution of the quantum circuit. Finally, we claim that for every cases we consider, at least for shallow circuits, there exists a positive correlation between the complexity of QUBO problems and information scrambling, as well as between the complexity and entanglement.

In what follows, we first briefly review the background of QAOA and tripartite information and introduce our implementation in Sec.~\ref{sec:Preliminaries}, and then compare information scrambling and entanglement entropy with complexity of problems in Sec.~\ref{sec:Scrambling-in-QAOA} and Sec.~\ref{sec:entanglement-in-QAOA} respectively. At last, we discuss our results and provide future lines of research in Sec.~\ref{sec:Conclusion}.

\section{Preliminaries}\label{sec:Preliminaries}

In this section, we briefly review the basic concepts on QAOA and tripartite mutual information describing information scrambling. After that, we introduce our basic settings in this work.

\subsection{The quantum approximate optimization algorithm}\label{subsec:QAOA}

As we discussed in Sec.~\ref{sec:introduction}, QAOA is a quantum algorithm aiming to solve combinatorial optimization problems, which was first brought by Farhi, \emph{et al.}~\cite{Farhi-1411}. Usually, a combinatorial optimization problem can be represented by a QUBO model, where we can associate a weighted graph $G=\left(V,E\right)$ with nodes $V=\left\{ 1,2,\cdots,N\right\}$ and edges $E=\left\{\left(i,j\right),i\neq j\right\}$. In this graph, the weight of the edge $\left\{i,j\right\}$ is denoted by $w_{ij}$, and the weight of the node $k$ is denoted by $h_k$. Connection property of the graph can be characterized by density of the graph:
\begin{equation}
D=\frac{2E}{N(N-1)},\label{eq:density}
\end{equation}
where $E$ is the number of edges and $N(N-1)/2$ is the maximal potential connections. The graph can be mapped into an Ising model in quantum many-body physics with $V$ represents independent terms $Z_{i}$ and $E$ represents interaction terms $Z_{i}Z_{j}$. Therefore, the most general form of an Ising Hamiltonian reads
\begin{equation}
H_{C}=\mathop{\sum_{i,j=1}^{N}w_{ij}Z_{i}Z_{j}+\sum_{k=1}^{N}h_{k}Z_{k}\,},\label{eq:H_C}
\end{equation}

\noindent where $Z_{i}$ are $\sigma_{z}$ operators with eigenvalues $\pm1$, $w_{ij}$ and $h_{k}$ are coupling strengths of the interaction and independent terms. The target of QAOA is to achieve the minimal quantum mean value $\left\langle H_{C}\right\rangle$ via optimized parameters. Here, the QAOA circuit with $p$-layer and $2p$ parameters is as follows
\begin{equation}
\left|\vec{\beta},\vec{\gamma}\right\rangle =U_{B}\left(\beta_{p}\right)U_{C}\left(\gamma_{p}\right)\cdots U_{B}\left(\beta_{1}\right)U_{C}\left(\gamma_{1}\right)\mathrm{H}^{\otimes N}\left|0\right\rangle ^{\otimes N}\,,\label{eq:QAOA-cir}
\end{equation}

\noindent where $U_{B}\left(\beta_{i}\right)=e^{-i\beta_{i}H_{B}}$, $H_{B}=\sum_{i=1}^{N}X_{i}$, $U_{C}\left(\gamma_{i}\right)=e^{-i\gamma_{i}H_{C}}$, and $\mathrm{H}$ means Hardmard gate. The $2p$ parameters $\vec{\beta}=\left(\beta_{1},\beta_{2},\cdots\beta_{p}\right)$ and $\vec{\gamma}=\left(\gamma_{1},\gamma_{2},\cdots\gamma_{p}\right)$ are optimized by classical optimization algorithm, after that they are applied to the parameterized quantum circuit Eq.~$\left(\ref{eq:QAOA-cir}\right)$ to obtain a smaller mean value 
$\left\langle H_{C}\right\rangle $. When QAOA finds the minimum of $\left\langle H_{C}\right\rangle $,
it outputs a set of optimal parameters $\left\{ \vec{\beta}_{opt},\vec{\gamma}_{opt}\right\}$ at the same time, here state $\left|\vec{\beta}_{opt},\vec{\gamma}_{opt}\right\rangle $ is the ground state of Eq.~$\left(\ref{eq:H_C}\right)$.

We note that $\left\{ \vec{\beta}_{opt},\vec{\gamma}_{opt}\right\} $ has multiple solutions for a certain $H_{C}$ as one choose different initial parameters $\left\{ \vec{\beta}_{0},\vec{\gamma}_{0}\right\} $. The choice of 
$\left\{\vec{\beta}_{0},\vec{\gamma}_{0}\right\} $ may result in seeking $\left\{ \vec{\beta}_{opt},\vec{\gamma}_{opt}\right\} $
of the problem easier or harder. Thus, to characterize properties of QAOA circuits solving the problem $H_{C}$ with less bias, we should turn to statistical average values of numerous samplings with random initial parameters. 

\subsection{Information scrambling and entanglement}\label{subsec:QI-intro}

The ``butterfly effect'' in quantum systems, where localized quantum information spreads to degrees of freedom of the entire system as a result of a small disturbance, is referred to as information scrambling. Hosur, \emph{et al. }use information scrambling to characterize the ability of a channel to process quantum information, they quantify scrambling as tripartite mutual information, which can be understood as entanglement between input and output of the channel~\cite{Hosur-jhep-2016}. For any unitary channel $\hat{U}$ in a $N$-qubit Hilbert space, it can be decomposed to
\begin{equation}
\hat{U}=\sum_{i,j=1}^{2^{N}}U_{ij}\left|i\right\rangle \left\langle j\right|\,,\label{eq:unitary}
\end{equation}

\noindent where $\left\{ \left|i\right\rangle \right\} $is the basis of $2^{N}$ Hilbert space. Denoting 
$\left\{ \left|i\right\rangle \right\}$
as the basis of $N$-qubit input legs and $\left\{ \left|j\right\rangle \right\}$
as the basis of $N$-qubit output legs, the unitary operator $\hat{U}$ can be mapped to a state $\left|U\right\rangle$ in the form of
\begin{equation}
\left|U\right\rangle =\sum_{i,j=1}^{2^{N}}\frac{U_{ij}}{\sqrt{2^{n}}}\left|i\right\rangle \left|j\right\rangle \,,\label{eq:operator-state}
\end{equation}

\noindent which represents an entanglement state of input qubits and output qubits. Since $\left|U\right\rangle $ is a pure entangled
state, we can define entanglement entropy between input and output states. Usually we divide the input part into subsystems $A$ and
$B$, and output part into $C$ and $D$, then entanglement entropy of $AC$ reads
\begin{equation}
S_{AC}=-\mathrm{Tr}\left(\rho_{AC}\mathrm{log}_{2}\rho_{AC}\right)\,,\label{eq:entropy-AC}
\end{equation}

\noindent where $\rho_{AC}=\mathrm{Tr}_{BD}\left(\left|U\right\rangle \left\langle U\right|\right)$.
Similarly, mutual information between $A$ and $C$ can be obtained as
\begin{equation}
I\left(A:C\right)=S_{A}+S_{C}-S_{AC}\,,\label{eq:mutual-info}
\end{equation}

\noindent which can be understood as how much information from input subsystem $A$ spread into output subsystem $C$ via the channel $\hat{U}$. Considering different conditions the information spreading from $A$ to $C$ and $D$, tripartite mutual information is introduced with the form
\begin{equation}
I_{3}\left(A:C:D\right)=I\left(A:C\right)+I\left(A:D\right)-I\left(A:CD\right)\,,\label{eq:tri-info}
\end{equation}

\noindent which is well-defined to characterize information scrambling. If there is no scrambling, all information of $A$ will be found in $C$ or $D$ (suppose $\left|A\right|<\left|C\right|,\left|D\right|$), either $I\left(A:C\right)$ or $I\left(A:D\right)$ equals to $I\left(A:CD\right)$ and another one equals to zero, then Eq.~$\left(\ref{eq:tri-info}\right)$ turns to be zero. If information of $A$ is fully scrambled to the whole output system, we cannot recover the information only by accessing $C$ or $D$, which means $I\left(A:C\right)=I\left(A:D\right)\sim0$, then we obtain $I_{3}\left(A:C:D\right)\sim-I\left(A:CD\right)\sim-2\left|A\right|$.

Tripartite mutual information is also defined as topological entanglement entropy in topological quantum field theory~\cite{Kitaev-prl-2006}, which is topologically invariant and represents a universal characterization of the many-body entanglement in the ground state of a topologically ordered two-dimensional system. 

\subsection{Implementation}\label{subsec:Implementation}

Here we introduce some settings throughout the paper. For QAOA circuit, we use the package of \emph{Qcover} to perform our simulations~\cite{Zhuang-2112}, the classical optimization algorithm we choose is Constrained Optimization BY Linear Approximation (COBYLA). The number of qubits of problems we consider is all $N=7$.

\begin{figure}[t]
\centering
\includegraphics[width=1\columnwidth]{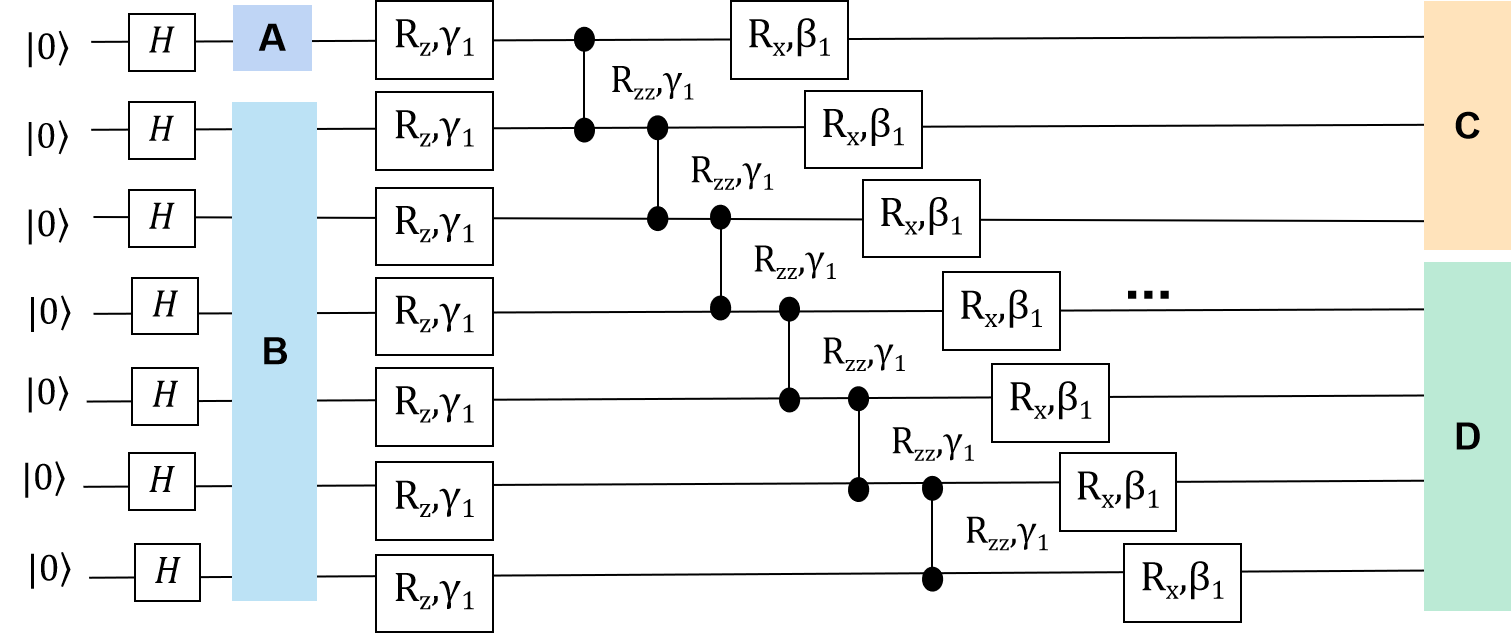}
\caption{\label{fig:QAOA} One layer of QAOA circuit and our choice of subsystems $ABCD$.}
\end{figure}

By taking the QAOA circuits as unitary channels to study information scrambling, we build up subsystems of $A$($\left|A\right|=1$), $B$($\left|B\right|=6$), $C$($\left|C\right|=3$) and $D$($\left|D\right|=4$) in Fig.~\ref{fig:QAOA}. Denoting that there are $245$ combinations for $ABCD$ with the same number of qubits, since QAOA circuits have no natural choice of these subsystems. We have tried several other combinations and find even $I_{3}\left(A:C:D\right)$ are different in values, the tendency in our results is invariant. Because tripartite information is a universal quantity describing scrambling ability of a channel, one choice is enough to reflect properties of the circuit we work on.

To study entanglement properties in detail, we consider von Neumann entanglement entropy of one qubit in the QAOA circuit. Since entanglement entropy
of one random qubit cannot imply the full structure of the entanglement properties, we compute average values of Von-Neumann entropy for every
qubits in the form of 
\begin{equation}
\bar{S}=\frac{-\sum_{i=1}^{N}\mathrm{Tr}\left(\rho_{i}\mathrm{log}_{2}\rho_{i}\right)}{N}\;,\label{eq:avg-entropy}
\end{equation}
where $\rho_{i}=\mathrm{Tr}_{\neg i}\left(\left|\vec{\beta}_{opt},\vec{\gamma}_{opt}\right\rangle \left\langle \vec{\beta}_{opt},\vec{\gamma}_{opt}\right|\right)$
is the reduced density matrix after tracing out information of all qubits except the $i$th qubit from final state computed by QAOA.

Finally, we come to a concrete definition of complexity of QUBO problems. We have introduced the attributes relevant to complexity in Eq. $\left(\ref{eq:def-complexity}\right)$ according to Ref.~\cite{QUBO_2022}, and now we need a corresponding function deduced from QAOA computing results to reflect this complexity. For a certain computing result $\left|\vec{\beta}_{opt},\vec{\gamma}_{opt}\right\rangle$, we take the distance between quantum mean value $\left\langle H_{C}\right\rangle $
and exact solution of the problem $E_{C}$ to be~\cite{Zhouprx2020,Zhang-npj-2022} 
\begin{equation}
\alpha=\frac{\left\langle H_{C}\right\rangle }{E_{C}}\;.\label{eq:alpha}
\end{equation}
The value was first introduced to benchmark the accuracy of approximate optimization algorithms~\cite{Hastad-acm-2001,Sakai-dam-2003}. As the initial parameters of QAOA are chosen randomly, if we compute the same problem for different times, the final parameters can be distinct from each other, which lead to different $\alpha$ values. We can only say, a harder problem has less probability for QAOA to achieve a high $\alpha$ value. Thus, in this work we characterize complexity of a QUBO problem as a statistical probability when the $\alpha$ value is larger than a certain bound we expect in a number of samplings. That is, if we sample a QUBO problem $n$ times and among them there are $s$ samples whose $\alpha$ values larger than the bound we set, then the complexity of the problem is a rate
\begin{equation}
R=\frac{s}{n}\;,\label{eq:complexity}
\end{equation}
which is also named as ``success rate''~\cite{Valle-pra-2021}. In ideal circumstances, this rate should be negatively correlated with complexity $\mathcal{C}$ as
\begin{equation}
\mathcal{C} \propto \frac{1}{R} \;.\label{eq:prop-complex}
\end{equation}
However, some problems, such as local minima problem caused by gradient based optimization algorithm~\cite{Bittel-prl-2021}, in QAOA may suppress $R$ and break the relation, which will be discussed in Sec.~\ref{sec:Scrambling-in-QAOA}.

\section{Scrambling in QAOA circuit versus Complexity of the problems}\label{sec:Scrambling-in-QAOA}

In this section, we investigate quantum information scrambling in QAOA circuits. The circuits are designed specially for solving QUBO problems with non-degenerate solutions, whose corresponding Ising Hamiltonians $H_C$ have separable ground states. As we have introduced in Sec.~\ref{subsec:QI-intro}, we take the circuits as unitary channels and quantify scrambling ability of the channels by tripartite information $I_3$. Our aim is to find some correlation between complexity of the problems and tripartite information.

\subsection{Comparison between different kinds of graphs}\label{subsec:Comparison-between-kinds}

As a start of this subsection, we first go back to the function of complexity Eq.~$\left(\ref{eq:def-complexity}\right)$. As we have fixed the degeneracy  of QUBO solutions and basic settings of QAOA, the computational complexity of QUBO problems should only relate to two factors: density of the graph $D$ and weights of the graph $\left\{w_{ij},h_k\right\}$. In order to present a clear comparison, in the following we choose three typical kinds of graphs: 

1.~Linear graphs with their corresponding Ising Hamiltonians in the form of $Z-ZZ$.

2.~Complete graphs with their corresponding Ising Hamiltonians in the form of $Z-ZZ$.

3.~Linear graphs with their corresponding Ising Hamiltonians in the form of $Z+ZZ$. 

\noindent Among the three cases, the $1$st and $2$nd cases have fixed graph weights but differ in graph density, whereas the $1$st and $3$rd cases have fixed graph density but differ in weights. According to Eq.~$\left(\ref{eq:prop-complex}\right)$, we expect if we fix one factor and change the other, the success rate $R$ will evolve monotonically.

\begin{figure}[htbp]
\centering
\includegraphics[width=1\columnwidth]{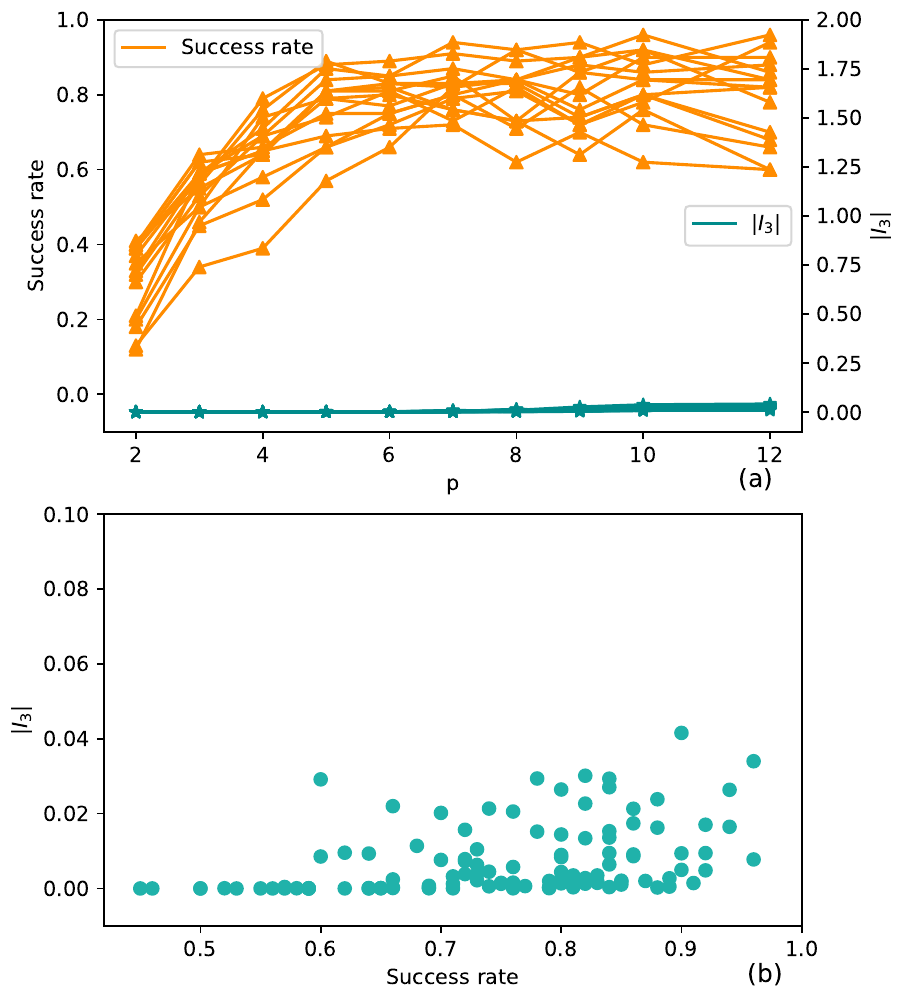}
\caption{\label{fig:Z-ZZ} The comparison between success rates to solve $\left\langle H_{linear}\right\rangle _{min}$ and information scrambling in the circuits with number of qubits $N=7$ for the \emph{linear} model of $Z-ZZ$-type Hamiltonians. Here $w\in\left[0.5,1.5\right]$, we randomly choose $15$ different weights. (a). In this figure, each point represents a $p$-layer QAOA circuit to compute $\left\langle H_{linear}\right\rangle _{min}$ with one randomly chosen $w$. We loop the QAOA circuit $10^{2}$ times, use Eq.~$\left(\ref{eq:complexity}\right)$ to obtain a success rate (orange point), and take an average value of $I_3$ (dark green point). We keep using these settings in all following $R-p$ and $I_3-p$ figures. The orange lines are success rates of QAOA with different layers, where we set $\alpha\protect\geqslant 0.996$, and entanglement entropy of $C$ and $D$ in final state $\protect\leqslant 0.02$. There are $15$ lines in the figure, representing $15$ problems solved by QAOA with different weights $w$. The dark green lines are average values of $\left|I_{3}\right|$ in these QAOA circuits corresponding to the orange lines. (b). Using the data from (a), we show a correlation plot for $R$ and $\left|I_{3}\right|$ from the same circuit (same $p$ and $w$).}
\end{figure}

\emph{1.~Linear graph with $Z-ZZ$-type Hamiltonian:} This kind of Ising Hamiltonians have only neighborhood interactions, which have the form of 
\begin{equation}
H_{linear}=+w\sum_{i=1}^{N}Z_{i}-\sum_{i=1}^{N-1}Z_{i}Z_{i+1}\,,\label{eq:H-line-exact}
\end{equation}

\noindent where $w$ is a weight parameter with $w\in\left[0.5,1.5\right]$. Since all QUBO problems corresponding to this kind of graphs have non-degenerate solutions, their Ising Hamiltonians share the same ground states $\left|1\right\rangle ^{\otimes N}$. As is shown in Fig.~\ref{fig:Z-ZZ}(a), success rate $R$ increases as the number of layers in QAOA circuit increases, and then becomes saturated after $p\sim6$. From Fig.~\ref{fig:Z-ZZ}(b) we note that even $I_{3}$ is very small, the positive correlation between $\left|I_{3}\right|$ and success rates is still visible.

To quantify the correlation, we introduce Pearson correlation coefficient here. Generally, Given two data sets $X$ and $Y$, the correlation between them can be characterized by 
\begin{equation}
\rho_{X Y}=\frac{\sigma_{X Y}}{\sigma_{X}\sigma_{Y}}\,,\label{eq:correlator}
\end{equation}
where $\sigma_{X Y}$ is their covariance, $\sigma_X$ and $\sigma_Y$ are their own variances. If two data sets have positive correlation, Pearson correlation coefficient takes the range of $(0,1]$. Now, by taking the data of $\left|I_{3}\right|$ and $R$ in Fig.~\ref{fig:Z-ZZ}(b), we can obtain the Pearson correlation coefficient $\rho_{I_3 R} \approx 0.4297$, which indicates an obvious positive correlation.

\noindent 

\begin{figure}[t]
\centering
\includegraphics[width=1\columnwidth]{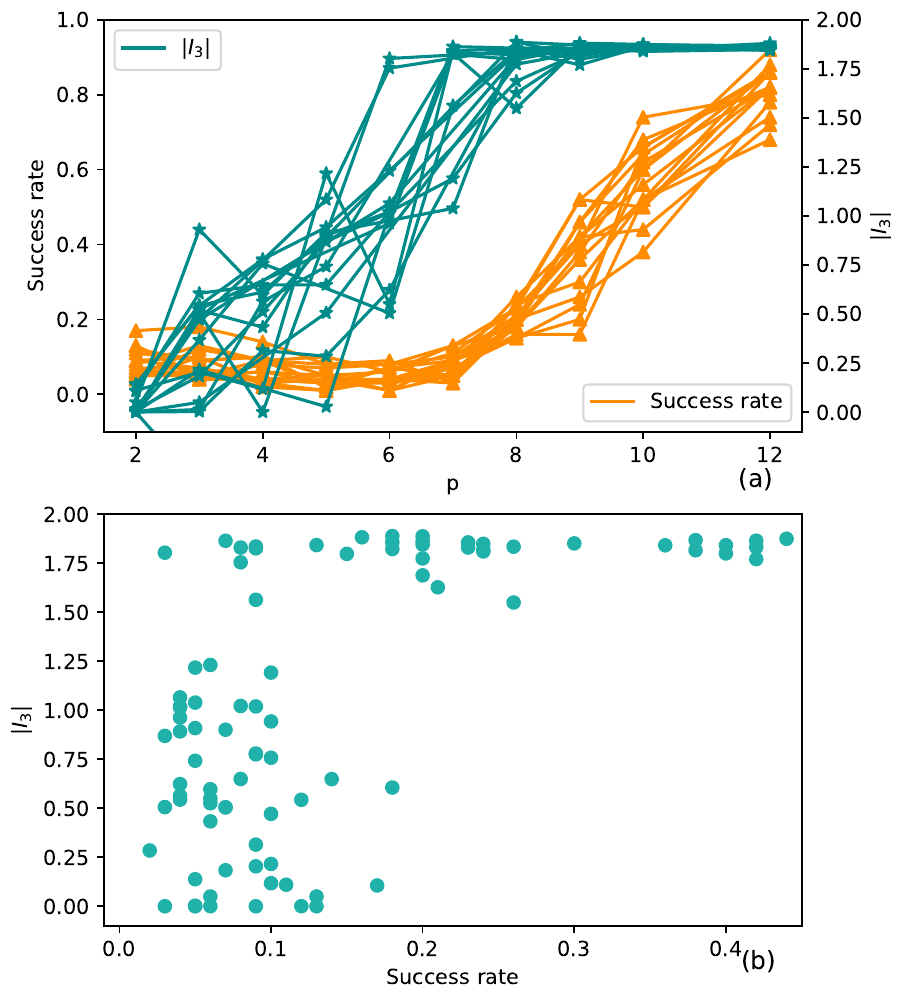}
\caption{\label{fig:Z-ZZ-complete} The comparison between success rates for $p$-layer QAOA to compute 
$\left\langle H_{complete}\right\rangle _{min}$ and information scrambling in the circuit for different layers with number of qubits $N=7$ for the \emph{complete} model with $Z-ZZ$-type Hamiltonians. Here $w\in\left[0.5,1.5\right]$, we randomly choose $15$ $w$s, and we take $10^{2}$ loops as well. (a). The orange lines are success rates of QAOA of different layers to compute $\left\langle H_{complete}\right\rangle _{min}$ with different $w$s, with the condition $\alpha\protect\geqslant0.996$, and entanglement entropy of $C$ and $D$ in final state $\protect\leqslant 0.02$. The dark green lines are $\left|I_{3}\right|$ in these QAOA circuits corresponding to the orange lines. (b). Using the data from (a), we show a correlation plot for $R$ and $\left|I_{3}\right|$ from the same circuit.}
\end{figure}

\emph{2.~Complete graph with $Z-ZZ$-type Hamiltonian:} For complete graphs, the corresponding Ising Hamiltonians are
\begin{equation}
H_{complete}=+w\sum_{i=1}^{N}Z_{i}-\sum_{i,j=1,i\neq j}^{N}Z_{i}Z_{j}\,,\label{eq:H-complete-exact}
\end{equation}
where $w\in\left[0.5,1.5\right]$ and each pair of nodes is connected by an edge. See Fig.~\ref{fig:Z-ZZ-complete}(a) we can find that, looking for the
minimal mean value of Eq.~$\left(\ref{eq:H-complete-exact}\right)$ is much harder than Eq.~$\left(\ref{eq:H-line-exact}\right)$, since the
success rates are much lower compared with the linear case especially for small layers. In this case, Eq.~$\left(\ref{eq:H-complete-exact}\right)$ have the same range of weights as Eq.~$\left(\ref{eq:H-line-exact}\right)$, but the graph density is much larger than Eq.~$\left(\ref{eq:H-line-exact}\right)$. It seems when we fix the weights, more connections make the QUBO problems harder to solve. About this issue, more details will be discussed in the next subsection. The success rates grow fast after $p\sim8$, tripartite information also grows to about $-1.86$ simultaneously. From Sec.~\ref{subsec:QI-intro} we have known the maximal tripartite information $I_{3}(A:C:D)$ is $-\left|A\right|\sim-2$, this means information encoded in the initial state $\left|+\right\rangle ^{\otimes N}$ should be sufficiently scrambled to reach the exact final state $\left|1\right\rangle ^{\otimes N}$ in the computing process. 

\noindent 

Moreover, a positive correlation is also presented in Fig.~\ref{fig:Z-ZZ-complete}(b). Using Eq.~$\left(\ref{eq:correlator}\right)$ we know Pearson correlation coefficient in this case is $\rho_{I_3 R} \approx 0.6581$. The correlation between tripartite information $I_3$ and success rate $R$ seems stronger than the above linear graph case. This phenomenon can be observed in Fig.~\ref{fig:Z-ZZ} and Fig.~\ref{fig:Z-ZZ-complete}, as the layer $p$ goes on, $R$ stops increasing at some value in Fig.~\ref{fig:Z-ZZ}(a) whereas $R$ in Fig.~\ref{fig:Z-ZZ-complete}(a) keeps increasing. Ideally, in a QAOA circuit, more layers indicate a more optimal result, and therefore a larger $R$ value. However, the classical optimization algorithm of QAOA is gradient based, when the function is complex and has many parameters, the minima evaluated by gradients may be far from the global minimum~\cite{Bittel-prl-2021}. This problem is an obstacle of all VQAs. For QAOA, more layers bring a larger optimal parameter space, which may leads to more local minima and suppress the value of $R$.


\emph{3. Linear graph with $Z+ZZ$-type Hamiltonian:} Now we study linear graphs with very different weights from Eq.~$\left(\ref{eq:H-line-exact}\right)$, namely
\begin{equation}
H^{\prime}_{linear}=+\sum_{i=1}^{N}Z_{i}+v\sum_{i,i+1}^{N-1}Z_{i}Z_{i+1}\,,\label{eq:H-line-hard}
\end{equation}
where $v\in\left[2,3\right]$ is the weight of edges $ZZ$. Unlike above cases, it is not easy to find the ground state at first glance. Among all $128$ basis of the $7$-qubit system, we find the product state $\left|1010101\right\rangle$ makes Eq.~$\left(\ref{eq:H-line-hard}\right)$ take the minimal mean value. 

\noindent 
\begin{figure}[tbp]
\centering
\includegraphics[width=1\columnwidth]{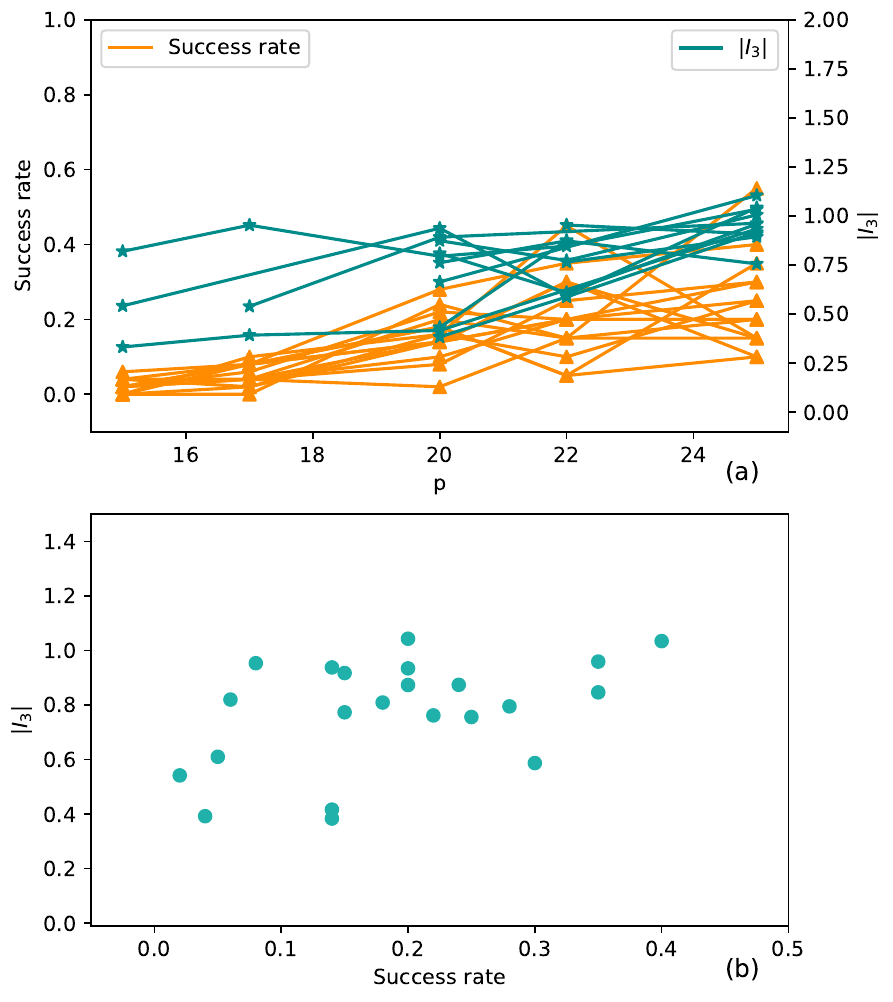}
\caption{\label{fig:Z+ZZ-linear}The comparison between success rates for $p$-layer QAOA to compute $\left \langle H^{\prime}_{linear}\right\rangle_{min}$ and information scrambling in the circuit for different layers with number of qubits $N=7$ for the \emph{linear} model with $Z+ZZ$-type Hamiltonians. Here $v\in\left[2,3\right]$, we randomly choose $15$ $v$s, and we take $10^{2}$ loops as well. (a). The orange lines are success rates of QAOA of different layers to compute $\left\langle H^{\prime}_{linear}\right\rangle _{min}$ with different $v$s, with the condition $\alpha\protect\geqslant 0.96$, and entanglement entropy of $C$ and $D$ in final state $\protect\leqslant 0.25$. The dark green lines are $\left|I_{3}\right|$ in these QAOA circuits corresponding to the orange lines. (b). Using the data from (a), we show a correlation plot for $R$ and $\left|I_{3}\right|$ from the same circuit.}
\end{figure}

As is shown in Fig.~\ref{fig:Z+ZZ-linear}, success rate (we set $\alpha>0.96$) are $\sim0$ before $p=15$, even when $p>15$ the probabilities are much smaller than above cases. Compared with Fig.~\ref{fig:Z-ZZ}, it is much harder to obtain a good result of ground state energy, and it seems more tripartite information is generated in the circuit. We note that, density of the graph $D$ of Eq.~$\left(\ref{eq:H-line-hard}\right)$ is the same as Eq.~$\left(\ref{eq:H-line-exact}\right)$, however different weights seem have a large contribution to complexity of the problems. With $D$ fixed, we propose the complexity $\mathcal{C}$ can be expressed as
\begin{equation}
\mathcal{C} \propto g\left(w_{ij},h_k\right)\, ,\label{eq:weights}
\end{equation}
\noindent where $g\left(w_{ij},h_k\right)$ only depends on weights of the graph.  
For the correlation between success rate and information scrambling, we get the Pearson correlation coefficient between tripartite information and success rate of Eq.~$\left(\ref{eq:H-line-hard}\right)$ is $\rho_{I_3 R} \approx 0.4461$. The correlation strength is similar to the first linear graph case.

\begin{figure*}[htbp]
\centering
\includegraphics[width=1.9\columnwidth]{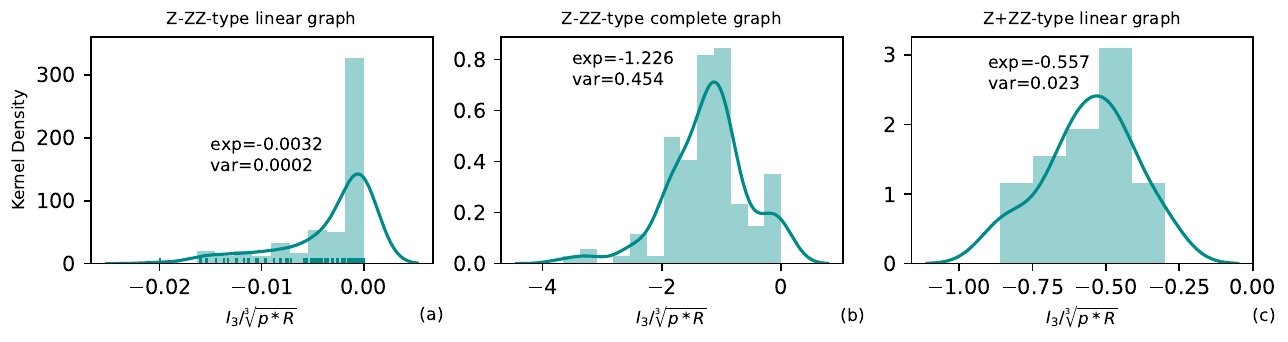}
\caption{\label{fig:ratio_ZZZ}\footnotesize Data distribution of $I_3/\sqrt[3]{p*R}$ for the three cases. Here in each cases we set $p \in [2,15]$, and take $10^2$ samplings for each $p$, to obtain success rate $R$ when $\alpha \geqslant 0.996$ and average tripartite information $I_3$. }
\end{figure*}

Having presented the positive correlation between tripartite information $I_3$ and success rate $R$ in the above three cases, let us estimate a possible relation between the two variables. As we have noticed, both $I_3$ and $R$ have upper bounds, and both of them increase fast at beginning then slow down when $p$ becomes larger. Therefore, we can try to fit the data presented in Fig.~\ref{fig:Z-ZZ}-\ref{fig:Z+ZZ-linear} via a root function like
\begin{equation}
I_3 \sim C_0 \sqrt[x]{p*R}\, ,\label{eq:relation}
\end{equation}
where $C_0$ is a constant, and power value $x$ differs in specific QUBO problems. For the QUBO problems studied in our work, we find the value of $x$ can be set to $x=3$. As is depicted in Fig.~\ref{fig:ratio_ZZZ}, a range of value for $C_0$ and a peak of the $C_0$ distribution can be found in each cases, which reflects a positive proportional relation between $I_3$ and $\sqrt[x]{p*R}$ approximately. Additionally, the circuit may reach ``entanglement barrier” when $p$ is large~\cite{Dupont-pra-2022,Chen-2205}, then $I_3$ won't increase with $p$. Therefore, this relation only holds for shallow QAOA circuits. When $p \gg 1$, both $I_3 \sim -2$ and $R \sim 1$, the ratio of $I_3$ and $\sqrt[x]{R}$ approximates to a constant.

To summarize, in this section we compare information scrambling in QAOA circuits with their corresponding success rates, which reflect the complexity of solving three different kinds of QUBO problems via QAOA. By comparing success rates $R$ in the three cases, we find it is sensitive to both factors of QUBO complexity $\mathcal{C}$, i.e. density of the graph $D$ and weights of the graph $\left\{w_{ij},h_k\right\}$. From those results, the positive correlation between tripartite information and complexity of the problems can be observed in the three kinds of graphs. Additionally, the forth case, complete graph with $Z+ZZ$-type Hamiltonians, should also be considered naturally. However, it's hard to find a non-degenerate solution, which cannot be mapped to a product ground state like above three cases. This could imply that for particular weights, it's difficult to avoid degeneracy in solutions, especially when the graph density is high.

\subsection{Comparison between different edges of graphs}\label{subsec:Comparison-between-edges}

In this subsection, we explore a more direct way to characterize the complexity of problems. If we fix weights of the graphs, the complexity Eq.~$\left(\ref{eq:def-complexity}\right)$ has only one variable, density of the graphs $D$. Since in this work we only discuss $N=7$, thus $D$ can be expressed by the number of edges $E$ according to Eq.~$\left(\ref{eq:density}\right)$. Then we can propose a relation between complexity and number of edges:
\begin{equation}
\mathcal{C} \propto E \;,\label{eq:prop-edge}
\end{equation}
when $\left\{w_{ij},h_k\right\}$ are fixed. To verify this correlation, next we focus on a set of graphs with fixed weights and different number of edges, whose corresponding Ising Hamiltonians are $Z-ZZ$-type:

\begin{subequations}
\begin{equation}
H_{1}=+\sum_{i=1}^{N}Z_{i}-\sum_{i=1}^{N-1}Z_{i}Z_{i+1}\;,
\end{equation}
\begin{equation}
H_{2}=+\sum_{i=1}^{N}Z_{i}-\sum_{i=1}^{N-1}Z_{i}Z_{i+1}-\sum_{i=1}^{N-2}Z_{i}Z_{i+2}\;,
\end{equation}
\begin{equation}
H_{3}=+\sum_{i=1}^{N}Z_{i}-\sum_{i=1}^{N-1}Z_{i}Z_{i+1}-\sum_{i=1}^{N-2}Z_{i}Z_{i+2}-\sum_{i=1}^{N-3}Z_{i}Z_{i+3}\;,
\end{equation}
\begin{align}
&H_{4}=+\sum_{i=1}^{N}Z_{i}-\sum_{i=1}^{N-1}Z_{i}Z_{i+1}-\sum_{i=1}^{N-2}Z_{i}Z_{i+2}-\sum_{i=1}^{N-3}Z_{i}Z_{i+3}\nonumber\\
&-\sum_{i=1}^{N-4}Z_{i}Z_{i+4}\;, 
\end{align}
\begin{equation}
H_{5}=+\sum_{i=1}^{N}Z_{i}-\sum_{i,j=1,i\neq j}^{N}Z_{i}Z_{j}\;.
\end{equation}
\label{eq:set-of-Hs}
\end{subequations}

\begin{figure}[htbp]
\centering
\includegraphics[width=0.9\columnwidth]{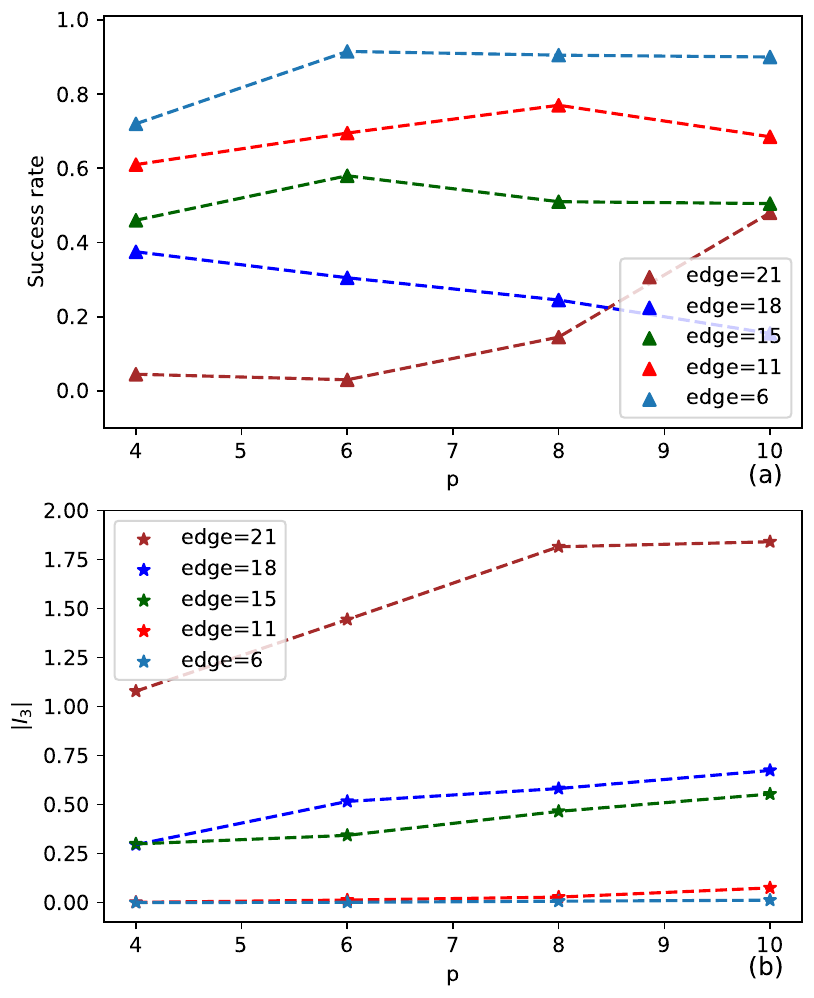}
\caption{\label{fig:edge-compare}\footnotesize The comparison between success rates for computing minimal mean values of 
Eq.~$\left(\ref{eq:set-of-Hs}\right)$ via $p$-layer QAOA and information scrambling in the circuit with number of qubits $N=7$, in $Z-ZZ$-type Hamiltonians with different edges $E$ and the same weights. (a). The success rates of QAOA of different layers when $\alpha\protect\geqslant 0.996$ and $10^{2}$ loops. (b). Average tripartite information of QAOA circuits corresponding to the success rates.}
\end{figure}

\noindent All the Hamiltonians in Eq.~$\left(\ref{eq:set-of-Hs}\right)$ have the ground state $\left|1\right\rangle ^{\otimes N}$. The relation between number of edges of these graphs $E$, success rates $R$ and tripartite information $I_3$ in QAOA circuit is depicted in Fig.~\ref{fig:edge-compare}. 

From Fig.~\ref{fig:edge-compare}(a) we see, generally speaking, the success rates $R$ is positively correlated with the number of edges $E$, which is in agreement with Eq.~$\left(\ref{eq:prop-edge}\right)$. However, as depth of the circuit goes deeper, the success rates seem stop increasing or even decrease, which is inconsistent with our ideal prediction. 

We have mentioned the local minima problem for QAOA in Sec.~\ref{subsec:Comparison-between-kinds}. A deeper circuit depth means a more complex process of variational optimization, and one more layer takes a pair of new parameters. Therefore, the algorithm faces more local minima and most of them have low $\alpha$ values. Another interesting phenomenon in Fig.~\ref{fig:edge-compare}(a) is, as the success rates of computing $\left\langle H_{1}\right\rangle _{min}$ to $\left\langle H_{4}\right\rangle _{min}$ stops increasing, computing $\left\langle H_{5}\right\rangle _{min}$ don't have to suffer from the problem. This may be explained that the complete graph is the most complex and needs more parameters for a good optimization, so when the layer $p$ is larger, relatively it has more possibility to reach a high success rate.

\begin{figure*}[htbp]
\centering
\includegraphics[width=1.9\columnwidth]{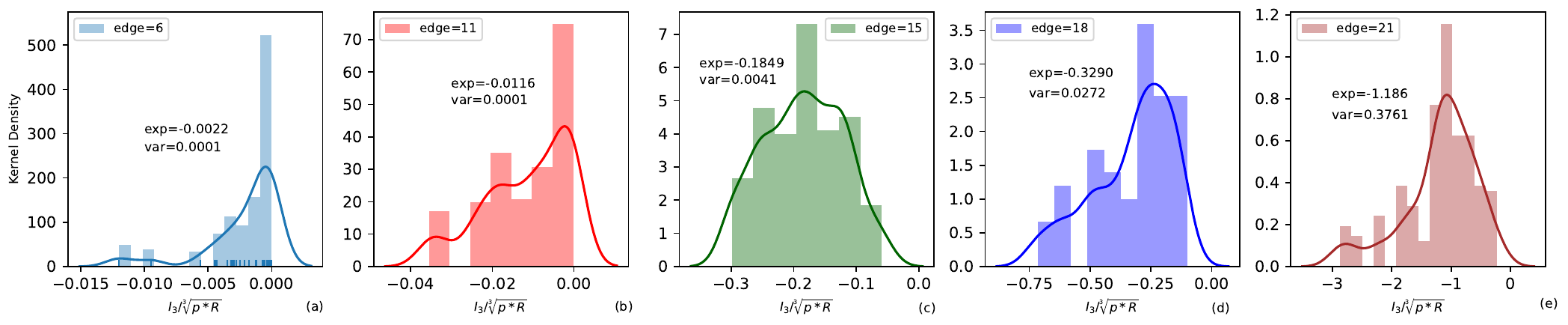}
\caption{\label{fig:ratio_Z-ZZ}\footnotesize Data distribution of $I_3/\sqrt[3]{p*R}$ for the QUBO problems in Eq.~$\left(\ref{eq:set-of-Hs}\right)$. Here in each cases we set $p=4,6,8,10$, and take $10^2$ samplings for each $p$, to obtain success rate $R$ when $\alpha \geqslant 0.996$ and average tripartite information $I_3$.}
\end{figure*}

From the distribution in Fig.~\ref{fig:ratio_Z-ZZ} we can find the relation Eq.~$\left(\ref{eq:relation}\right)$ holds for graphs in Eq.~$\left(\ref{eq:set-of-Hs}\right)$ as well. For a certain QUBO graph, one can estimate the value $C_0$ of Eq.~$\left(\ref{eq:relation}\right)$ to obtain an equation between $I_3$ and the cube root success rate $R$. We can also observe the expectation value of $C_0$ from Fig.~\ref{fig:ratio_Z-ZZ} increases as the number of edges $E$ goes on. It is because the success rate $R$ decreases when the QUBO problem becomes harder to solve.

Moreover, as the number of edges $E$ can reflect complexity of QUBO problems directly in Eq.~$\left(\ref{eq:prop-edge}\right)$, we wonder if there is also a positive correlation between number of edges $E$ and information scrambling $I_3$. The tripartite information in QAOA circuits versus different $E$ is depicted in Fig.~\ref{fig:edge-compare}(b), where we can observe a positive correlation between $E$ and $I_3$ as well. Here, the Pearson correlation coefficients of $E$ and $I_3$ are $0.8357, 0.8624, 0.8512$ and $0.8761$ for four different layers $p=4, 6, 8, 10$.

\section{Entanglement in QAOA circuit versus complexity of the problems}\label{sec:entanglement-in-QAOA}

\noindent 

\noindent 

As we have discussed in previous sections, information scrambling is a global feature of a quantum channel. Next, we turn to more details for entanglement properties in the circuits. We still take Eq.~$\left(\ref{eq:set-of-Hs}\right)$ as the set of our interested QUBO problems with non-degenerate solutions, and their corresponding Ising Hamiltonians have the ground state $\left|1\right\rangle ^{\otimes N}$. As have introduced in Sec.~\ref{sec:introduction}, since both the initial and final state are separable, entanglement is only generated during the computing process. Therefore in this section, we explore how to quantify the entanglement generated in a QAOA circuit, and its correlation with complexity of QUBO problems.

\begin{figure}[htbp]
\centering
\includegraphics[width=0.9\columnwidth]{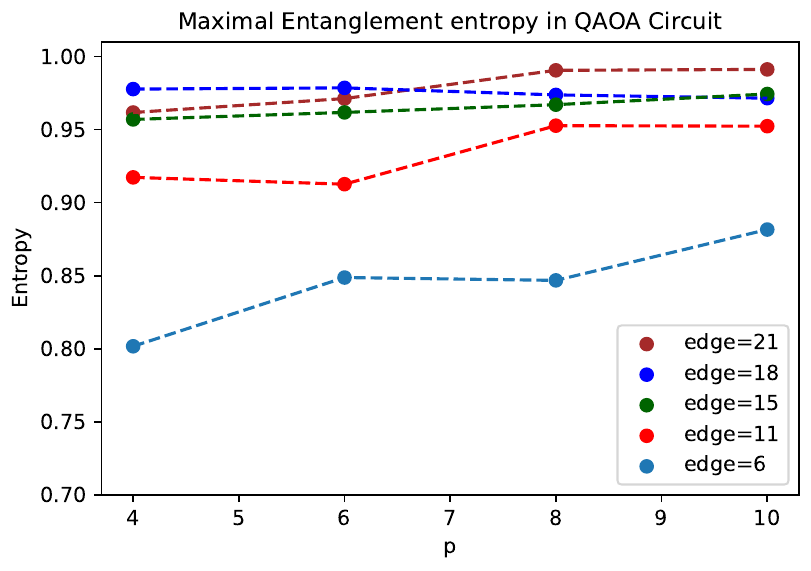}
\caption{\label{fig:one-qubit-entropy}\footnotesize One-qubit maximal entanglement entropy in $p$-layer QAOA circuits during the process of computing minimal mean values of Ising Hamiltonians in Eq.~$\left(\ref{eq:set-of-Hs}\right)$. Here we choose four different layers: $p=4,6,8,10$, and take an average of $10^{2}$ data. }
\end{figure}

\begin{figure}[htbp]
\centering
\includegraphics[width=1.0\columnwidth]{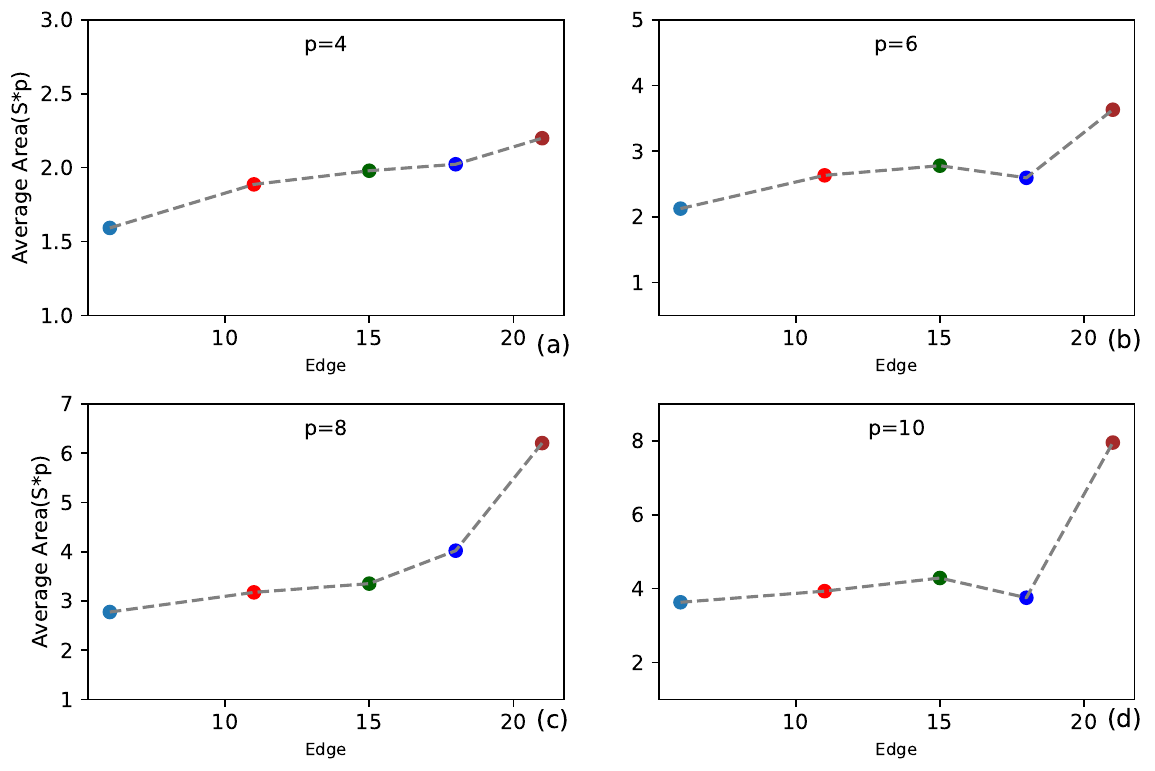}
\caption{\label{fig:entropy-edge}\footnotesize Average entanglement accumulation $\bar{\mathcal{A}}$ to reach $\alpha\geqslant 0.996$ during the process of computing minimal mean values of Hamiltonians in Eq.~$\left(\ref{eq:set-of-Hs}\right)$with four fixed layers $p=4,6,8,10$. We take an average of $50$ samplings for each case.}
\end{figure}

We first define a coarse-grained quantity for a QAOA circuit with a given layer $p$: 
\begin{equation}
S^{p}_{max} = \mathrm{max}_{l \leqslant p, l\in \mathbb{Z}}\left\{\bar{S}^{p}_{l}\right\}\, ,\label{eq:Smax}
\end{equation}
\noindent where $\bar{S}^{p}_{l}$ has the form of Eq.~$\left(\ref{eq:avg-entropy}\right)$, which is the average one-qubit entanglement entropy of every layer $l\leqslant p$ in the $p$-layer circuit. Eq.~$\left(\ref{eq:Smax}\right)$ means the maximum of average one-qubit entanglement entropy among all $\bar{S}^{p}_{l}$. After that, we present the maximal entanglement entropy generated in the circuits during the process of computing minimal mean values of Hamiltonians of Eq.~$\left(\ref{eq:set-of-Hs}\right)$ in Fig.~\ref{fig:one-qubit-entropy}. From this figure, we can see the tendency that the more number of edges $E$ the graph has, the more entanglement is generated in the circuit. Similar to Sec.~\ref{sec:Scrambling-in-QAOA}, we quantify the correlation via Pearson coefficients with number of edges $E$ and maximal one-qubit entanglement entropy $S^{p}_{max}$, they are $0.8982, 0.9446, 0.9033, 0.9255$ for layers $p=4, 6, 8, 10$ respectively.

To this point, a question raises naturally: What about the entanglement properties inside one layer? As previous studies of entanglement in QAOA circuit only take one ``layer'' as a unit, what happens inside one layer has not been investigated yet. According to the layer of QAOA circuit shown in Fig.~\ref{fig:QAOA}, the two-qubit gates represent the terms $e^{-i \gamma_i Z_j Z_k}$ in Eq.~$\left(\ref{eq:QAOA-cir}\right)$, all of them can generate entanglement. Therefore, in the following we step into each layer and focus on total entanglement accumulated in QAOA circuits.

\begin{figure}[htbp]
\centering
\includegraphics[width=0.85\columnwidth]{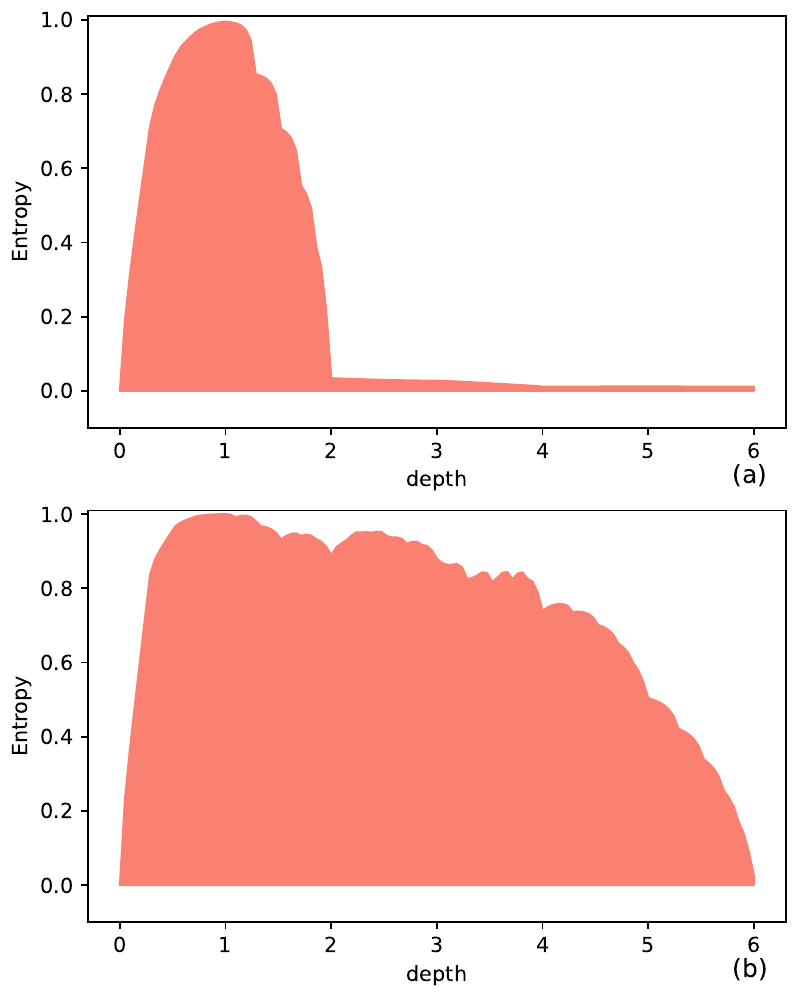}
\caption{\label{fig:examples}\footnotesize Two examples of $S-p$ plots for entanglement entropy accumulation when $p=6$, edges of $ZZ=21$. Here we calculate $\bar{S}$ after each two-qubit gate in the circuit, and the caption ``depth'' means the layers $l\leqslant p$ of one $p$-layer QAOA circuit. }
\end{figure}

We now introduce a proper characterization of entanglement accumulation in $p$-layer QAOA circuit in the form of
\begin{equation}
\mathcal{A} = Area\left(\bar{S}*p\right)\, ,\label{eq:area}
\end{equation}
which is the area of $S-p$ plot for average one-qubit entanglement entropy $\bar{S}$ generated during the evolution of $p$-layer QAOA circuit. Ideally, we expect the accumulation $\mathcal{A}$ is always positively correlated with number of edges $E$. However, due to the randomness of initial parameters $\left\{\vec{\beta}_0,\vec{\gamma}_0\right\}$, entanglement accumulation of the same layer in the same computing process can be very different from each other. For example, two different typical structures of entanglement accumulation in the process of computing $\left\langle H_{5}\right\rangle _{min}$ when $p=6$ is shown in Fig.~\ref{fig:examples}. The structure in Fig.~\ref{fig:examples}(a) only demands large entanglement at first two layers, whereas entanglement in Fig.~\ref{fig:examples}(b) is always large until the final step. Thus, to obtain a practical quantity describing entanglement accumulation, we still need a number of samplings and consider average values.

Here, we present the distribution of $\mathcal{A}$ for $50$ samplings per layer and per problem in Fig.~\ref{fig:dis}. We can find for harder problems, the ``tails'' of probability density become larger, which means the proportion of entanglement structures like Fig.~\ref{fig:examples}(b) increases, and indicates a requirement for more entanglement accumulation. Moreover, by taking average values of these samplings, we obtain a positive correlation between average areas of entanglement accumulation $\bar{\mathcal{A}}$ and number of edges $E$ of the graphs of Eq.~$\left(\ref{eq:set-of-Hs}\right)$, which is shown in Fig.~\ref{fig:entropy-edge}. 

\begin{figure}[htbp]
\centering
\includegraphics[width=1.0\columnwidth]{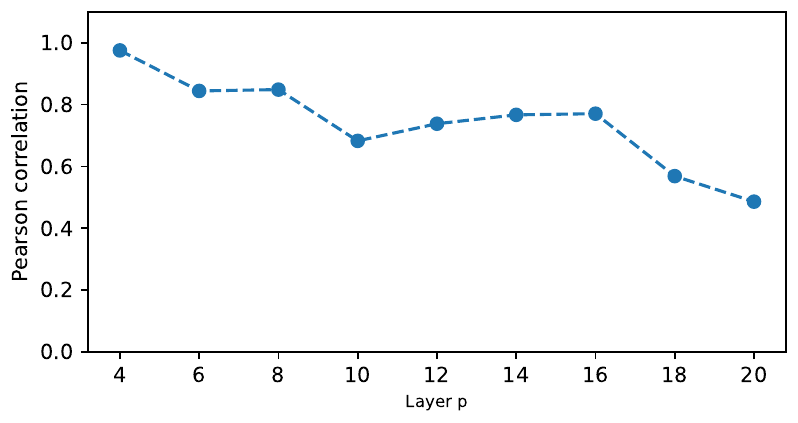}
\caption{\label{fig:corr_dec}\footnotesize The decrease of Pearson correlation coefficients between average entanglement accumulation $\bar{\mathcal{A}}$ and number of graph edges $E$ as the layer $p$ goes on.}
\end{figure}

However, as the layer $p$ increases, the effect of ``entanglement barrier'' cannot be neglected~\cite{Dupont-pra-2022,Chen-2205}, which can weaken the strength of positive correlation. For the correlation coefficients between average areas of entanglement accumulation $\bar{\mathcal{A}}$ and number of edges $E$, we can observe this decrease. As is shown in Fig.~\ref{fig:corr_dec}, from which we find that when $p\gg 20$, it can be harder to observe the positive correlation because the entanglement barrier arrives.

\noindent 

\section{Conclusions and outlook}\label{sec:Conclusion}

In this work, we have addressed the question of how much quantum resource is consumed for solving QUBO problems with different complexity in QAOA circuits. More precisely, we characterize quantum resource in QAOA circuits by two concepts in quantum information, information scrambling and entanglement, and characterize complexity of problems by success rates and number of edges of graphs. By taking QAOA circuits as unitary quantum channels, we use tripartite mutual information to quantify information scrambling. Next, we use the maximal entanglement entropy among every layers in QAOA circuits to present how much entanglement is needed in the whole computing process. After that, we study entanglement accumulation caused by two-qubit gates inside every layers in QAOA circuits, where we define an effective quantity to describe this process, namely an average area of entanglement entropy generated in the circuits and depth of the circuits.

Our main contribution is to present a positive correlation between complexity of QUBO problems and quantum resources including information scrambling and entanglement in QAOA circuits. Specially, we find the tripartite mutual information in QAOA circuits is positively proportional to the cube root of success rate approximately. Apart from this contribution, we clarify the factors related to the complexity of QUBO problems $\mathcal{C}$: graph weights $\left\{w_{ij},h_k\right\}$ and graph density $D$. Then, we use success rate $R$ to quantify the complexity in specific cases of our work. After that, we offer two quantities to describe how much entanglement, i.e. the maximal entanglement entropy Eq.~$\left(\ref{eq:Smax}\right)$ and the area of entanglement accumulation Eq.~$\left(\ref{eq:area}\right)$, is required in QAOA circuits, which can be useful for further applications to practical quantum circuits.

To give a deeper explanation to the positive correlation, it is noteworthy that our results uncover a map between computational complexity of QUBO models, and physical complexity of the unitary channels in corresponding QAOA circuits. More precisely, the computational complexity of a QUBO model is governed by its geometric structure, which is a mathematical property. This geometric structure, in turn, dictates the physical structure of the corresponding Ising Hamiltonian. Ultimately, this physical structure determines complexity of the circuit employed in QAOA. From the perspective of many-body physics, for an Ising Hamiltonian exhibiting increased complexity in its physical structure, achieving evolution to its ground state from the initial state of QAOA becomes significantly more complex. Therefore, the process requires more quantum resource. In this work, we numerically present the map between the two complexities in our results. For future works, we will explore the connections more analytically.

Moreover, several remarks and outlooks including scrambling and entanglement are discussed as follows.

In the part of information scrambling, as we have discussed in Sec.~\ref{sec:Scrambling-in-QAOA}, the circuits are more scrambled when the problem is more complex. However, there is a question: How to compare the complexity of case Eq.~$\left(\ref{eq:H-complete-exact}\right)$ and case Eq.~$\left(\ref{eq:H-line-hard}\right)$? Obviously the success rates of Eq.~$\left(\ref{eq:H-line-hard}\right)$ is much lower, but its tripartite information is smaller than that of Eq.~$\left(\ref{eq:H-complete-exact}\right)$. To this extent, we need further studies to well-define the relation between complexity $\mathcal{C}$ in a general context. 
 
It should be pointed out that the resource of information scrambling includes two aspects: entanglement scrambling and magic scrambling~\cite{Mi-science-2021,Garcia-2208}. Among these aspects, magic is said to be a complete quantum resource as it cannot be simulated by classical computers, and it is used to quantify complexity of fidelity estimation~\cite{Leone-pra-2023,Leone-2305}. Therefore, an interesting future work is to characterize how much magic generated in QAOA circuits, which can be an important direction for demonstrating quantum advantages beyond classical computation.

\begin{figure*}[htbp]
\centering
\includegraphics[width=1.9\columnwidth]{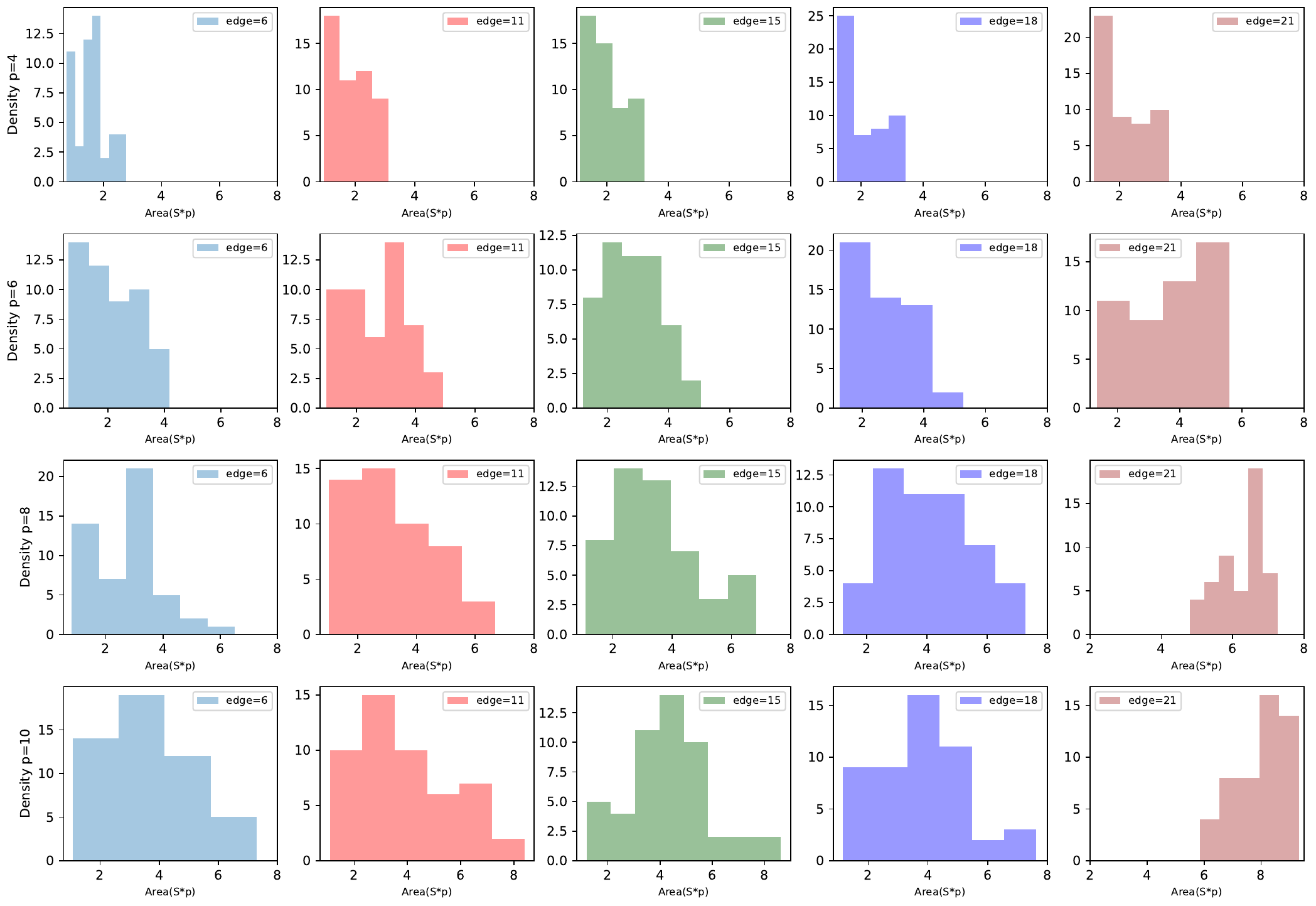}
\caption{\label{fig:dis}\footnotesize The distributions and probability density of areas of $S-p$ plots with different QAOA layers and problems for $50$ samplings. Figures from up to down are distributions of the same problems for layers $p=4, 6, 8, 10$. Figures from left to right are distributions of the same layers for the problems $\left\langle H_{1}\right\rangle _{min}$ to $\left\langle H_{5}\right\rangle _{min}$.}
\end{figure*}

In the part of entanglement, the average area of entanglement accumulation $\bar{\mathcal{A}}$ can be viewed as an overall entanglement resource required in the process of solving target problems, which can be a general quantity not only for a certain kind of problems like Eq.~$\left(\ref{eq:set-of-Hs}\right)$. We can apply this quantity to all QUBO problems with non-degenerate solutions, and generalizing the quantity to include problems with degenerate solutions is also a promising topic be discussed in the future. Moreover, as is shown in Fig.~\ref{fig:examples} and Fig.~\ref{fig:dis}, $\mathcal{A}$ for the same problem can be very different due to random choices of $\left\{ \beta_{0},\gamma_{0}\right\} $. It is interesting to consider the entanglement accumulation as a special feature of QAOA optimization process. Then, we can improve the algorithm by selecting the output that requires the least amount of entanglement accumulation among outputs with the same $\alpha$ value.

In summary, our results try to build a connection between the computational complexity of mathematical problems and physical quantum resources required for solving them in corresponding QAOA circuits. It should be noticed that our discussion is so far limited to small number of qubits and easy problems. In further studies, a generalization of our results may has a wide application in areas of not only quantum computation but also many-body physics, such as many-body localization~\cite{Abanin2019RMP}, quantum phase transitions~\cite{Dziarmaga-ap-2010}, and so on.

\section*{Acknowledgments}

We thank Ya-Nan Pu, Yunheng Ma and Yanwu Gu for useful discussions. This work is supported by Beijing Natural Science Foundation (No.Z220002) and National Natural Science Foundation of China (Grant No.12305010).

\section*{Data Availability Statement}

This manuscript has associated data generated from quantum software \emph{Qcover} developed by Beijing Academy of Quantum Information Sciences, with the GitHub link: \href{https://github.com/BAQIS-Quantum/Qcover/tree/main/Qcover/research}{https://github.com/BAQIS-Quantum/Qcover/tree/main/Qcover/research}. [Authors' comment: The datasets generated for the study, together with the code used for the analysis, are available from the corresponding author on request.]

\noindent 

\section*{}

\bibliography{quantum-information-QAOA}
\end{document}